\documentclass{article}\usepackage{jkas2}
\runningauthor{Mazzotta et al.}
\runningtitle{Heated Intracluster Gas and Radio Connections}
\beginpage{1}
\endpage{5}

\def\chandra    {{\em Chandra}\/}
\def\rxj1720    {{ RXJ1720.1+2638}\/}

\def\second      {{\prime\prime}}

\begin{document}

\font\twelvei = cmmi10 scaled\magstep1 
       \font\teni = cmmi10 \font\seveni = cmmi7
\font\mbf = cmmib10 scaled\magstep1
       \font\mbfs = cmmib10 \font\mbfss = cmmib10 scaled 833
\font\msybf = cmbsy10 scaled\magstep1
       \font\msybfs = cmbsy10 \font\msybfss = cmbsy10 scaled 833
\textfont1 = \twelvei
       \scriptfont1 = \twelvei \scriptscriptfont1 = \teni
       \def\mit{\fam1 }
\textfont9 = \mbf
       \scriptfont9 = \mbfs \scriptscriptfont9 = \mbfss
       \def\bmit{\fam9 }
\textfont10 = \msybf
       \scriptfont10 = \msybfs \scriptscriptfont10 = \msybfss
       \def\bmsy{\fam10 }

\def\etal{{\it et al.~}}
\def\eg{{\it e.g.,~}}
\def\ie{{\it i.e.,~}}
\def\lsim{\raise0.3ex\hbox{$<$}\kern-0.75em{\lower0.65ex\hbox{$\sim$}}}
\def\gsim{\raise0.3ex\hbox{$>$}\kern-0.75em{\lower0.65ex\hbox{$\sim$}}}

\title{Heated Intracluster Gas and Radio Connections: the Singular case of 
MKW~3s}

\author{Pasquale Mazzotta\altaffilmark{1,2}, 
Gianfranco Brunetti\altaffilmark{3},
Simona Giacintucci\altaffilmark{4},
Tiziana Venturi\altaffilmark{3},
Sandro Bardelli\altaffilmark{4}}
\address{
$^{1}\,${Dipartimento di Fisica, Universit\`a di Roma Tor Vergata,  
via della Ricerca Scientifica 1, I-00133 Roma, Italy}\\
{\it E-mail: mazzotta@roma2.infn.it}}
\address{$^{2}\,${Harvard-Smithsonian Center for Astrophysics, 60 Garden  
Street, Cambridge, MA02138, USA}}
\address{$^{3}\,${Istituto di Radioastronomia del CNR, Via P. Gobetti
    101, I-40129, Bologna, Italy}}
\address{$^{4}\,$INAF - Osservatorio Astronomico di Bologna, via Ranzani 1, I-40127 Bologna, Italy}

%


\address{\normalsize{\it (Received October 31, 2004; Accepted December 1,2004)}}
\abstract{Similarly to other  cluster of galaxies  previously classified as cooling
  flow systems,  the \chandra~ observation of MKW~3s reveals that  
this  object  has a  complex X-ray structure hosting both  a  X-ray
cavity and a X-ray filament.  Unlike the other clusters, however, the
temperature map of the core of MKW~3s shows the presence of extended
regions of gas heated above the radially averaged gas temperature at any radius.
 As the cluster does not show
evidences for  ongoing  major mergers Mazzotta et al. suggest a
connection between the heated gas and the activity of the central
AGN. Nevertheless, due to the lack of high quality radio maps,
this interpretation was controversial.
 In this paper we
present the results of two new radio observations of MKW~3s at 1.28~GHz
and 604~MHz  obtained at the GMRT. 
Together with the \chandra ~ observation and a separate VLA observation at 327~MHz from  Young, 
we show unequivocal  evidences for a close connection between the
heated gas region and the AGN activity and we briefly summarize possible implications.
}
\maketitle

\section {Introduction}
Although cooling flows (see Fabian 1994 for a review) is expected  to be the
natural state of cluster cores XMM and \chandra ~ observations show that
little of the gas cools below 1~keV (see e.g. Peterson et al. 2001, Kaastra et al. 2001, Tamura et al. 2001).
To explain the observed disagreement  with ``simple'' cooling flow
expectations, a number of authors propose different  physical
mechanisms able to provide enough heating to the gas 
to compensate the radiative losses (see e.g. Fabian et al. 2001). 
One of such mechanism is heating by an active galactic nucleus (AGN; e.g.
Binney  \& Tabor  1995).
Although  many \chandra ~ observations clearly show that AGN
may efficiently  inflate bubbles in the hot gas, to date it is not clear
if they may provide enough heat to quench colling flows (see
e.g. Birzan et al. 2004). 
In some cases it has been even raised the question if  bubbles inflated
by radio jet can even heat the gas at all. Earlier X-ray observations, in
fact, did not show any strong indication of heated gas in proximity of
the expanding bubbles (e.g. Allen et al. 2001),  contrary to what expected in jet-feed radio lobes
models that predict possible formations of shocks waves. In some cases,
as in Perseus, the data show that the coolest gas in the core lies immediately
around the radio lobes (Fabian et al. 2000). 
Nevertheless, evidences for the presence of weak  shock has began to
emerge recently with notably examples in Perseus  (Fabian et
al. 2003), Virgo (Forman et al. 2003), and Hydra A (Nulsen et al. 2004).

Before those, evidences for gas heating  possibly connected with the
activity of the central AGN were reported in MKW~3s by Mazzotta et
al. (2002). Despite of the  large extent of the heated gas
region,  its  connection with the central
AGN activity was a bit controversial due to the lack of high quality radio observations.
In this paper we present  the results of the analysis of two new radio
observations of MKW 3s at 1.28 GHz and
604 MHz obtained at the Giant Metrewave Radio  Telescope (GMRT, Pune, India).
By comparing them with the  \chandra ~ results of Mazzotta et al. 2002 
and making use of a 327~MHz radio map from  Young (2004) obtained at the
Very Large Array (VLA, Socorro, New Mexico, USA)  
we show clear evidences of a close connection  between the AGN activity
and the heated gas region in this cluster.

We use $H_0=70$~km~s$^{-1}$~kpc$^{-1}$, which implies a linear scale of
0.87~kpc per arcsec at the distance of MKW3s ($z=0.045$).


\begin {figure*}[t]
\vskip 0cm
\centerline{\centering \leavevmode
\epsfysize=1.\columnwidth 
\epsffile{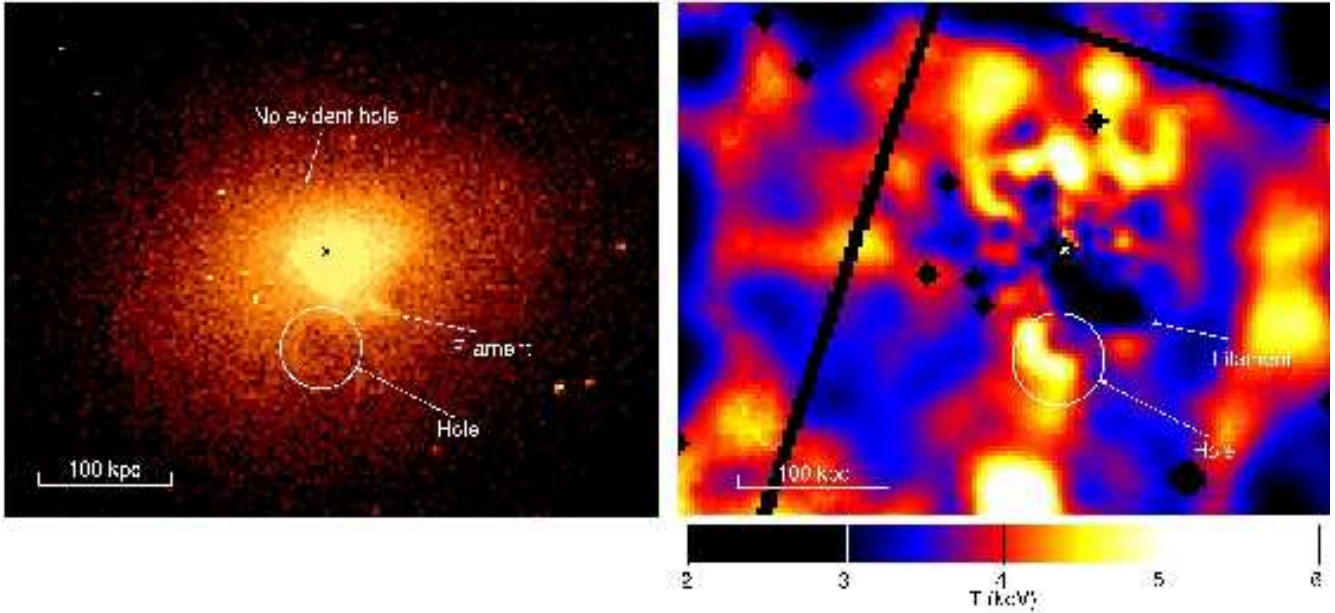}}
\label{fig:x-ray_t}
\caption{{\it left panel}) \chandra ~ image of the central $9.5$~arcmin$\times 7.5$~arcmin
($490$~kpc$\times 390$~kpc) region of 
 MKW3s from Mazzotta et al. (2002).  The X-ray image is obtained in the $0.5-8$~keV energy band
and each pixel corresponds to $4^\second \times 4^\second$. 
The  X symbol in the center and the arrow indicate the position of the 
 X-ray peak and the filament, respectively. The circle to the south of
 the X-ray peak indicate the position of the X-ray hole. We notice that
 this image does not clearly reveal the presence of a second X-ray hole
 opposite to the southern one. 
{\it Right panel})  Temperature map of  MKW3s from Mazzotta et
al. (2002).  All symbols are located in the same positions as in the
left panel.  The  statistical
  error in the temperature map is $<\pm 0.4$~keV at 68\% significance
  level ( $<\pm 0.8$~keV at 90\%). }
\vskip -0.5cm
\end{figure*}


\begin {figure*}[t]
\vskip 0cm
\centerline{\centering \leavevmode
\epsfysize=1.\columnwidth 
\epsffile{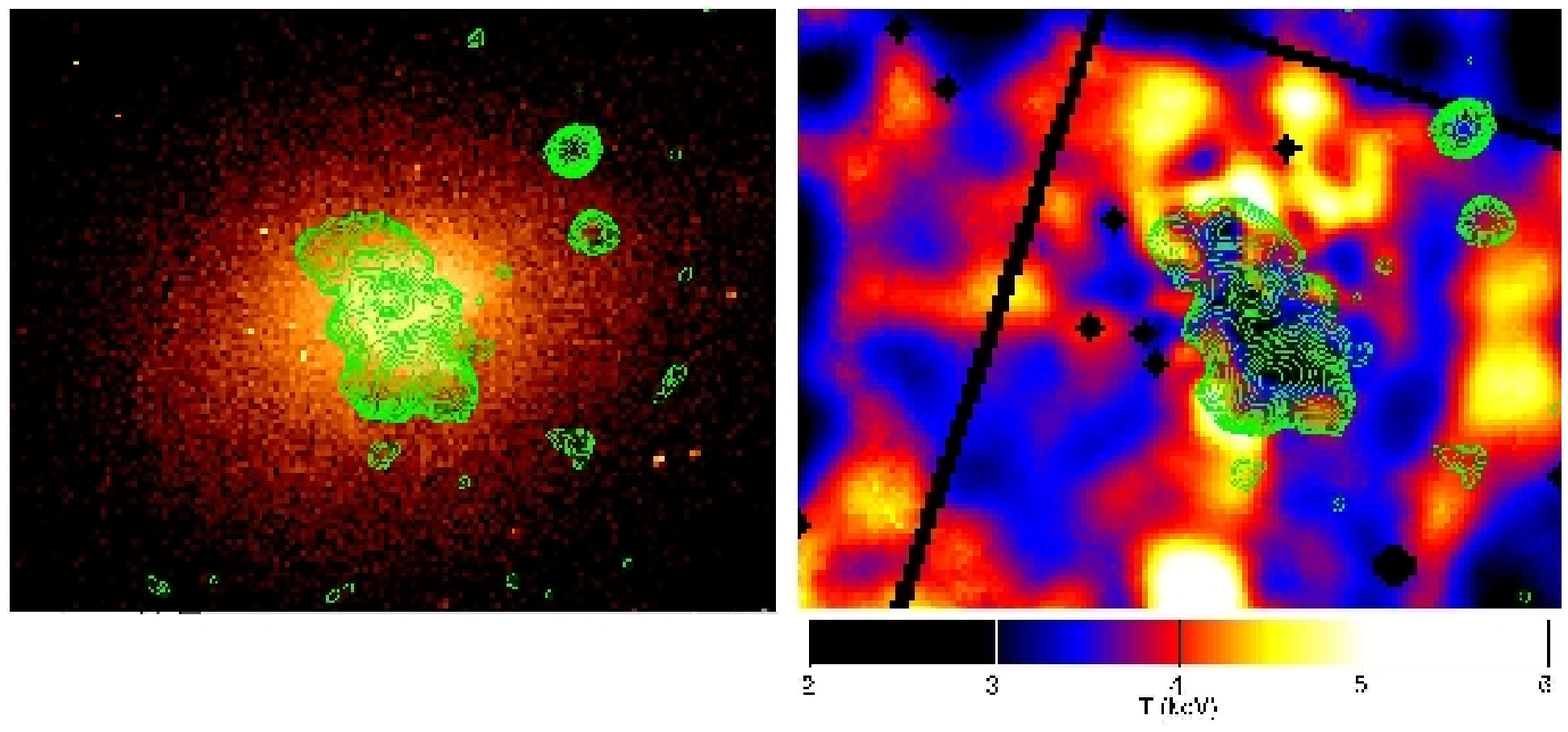}}
\label{fig:x-ray_t_radio}
\caption{ \chandra ~ image and temperature map of   
 MKW3s  as shown in Figure~1 overlaid to the~1.28 GHz radio contours obtained with the GMRT.
The resolution of the radio map is $16.8^\second \times 15.8^\second$.
Isoncontours levels are  equispaced  by a factor of $\sqrt{2}$ with the
lowest contour corresponding to 0.15 mJy.}
\vskip -0.5cm
\end{figure*}

\section{X-ray Observation}

MKW~3s was observed  on April 2000 in  ACIS-I with an exposure of
$\approx 57$~ksec.  A first analysis of these data
done by Mazzotta et al. (2002) revealed that the center of this
cluster is quite complex. Here we highlight only the most interesting
features and we refer to  Mazzotta et al. (2002) for details. 

 In the left panel of Figure~1
 we report the $[0.5-8]$~keV, background subtracted, vignetting corrected  \chandra ~
image of MKW~3s from Mazzotta et al. (2002).
The most interesting features  
are: i) a  surface brightness filament 
extending from the cluster center to the south-west; ii) a 
circular-like depression  to the south of the X-ray peak.

In the right panel of Figure~1 we  report 
 the temperature map from Mazzotta et al. (2002).
This image reveals that the temperature structure of the core of MKW~3s is even
more complex than its X-ray surface brightness.
Mazzotta et al. (2002) focused on the nature of  the extended 
hot regions $>5$~keV. 
  The most interesting aspect of these regions  is that their temperature   
is significantly higher than the radially averaged gas temperature at any radius (see
Figure~3 and  Figure~4 of Mazzotta et al. 2002).
As  both the X-ray surface brightness structure at larger
radii and the galaxy velocity distribution do not indicate any
significant ongoing major merger event (see e.g. Buote \& Tsai 1996,
Giradi et al. 1997),  Mazzotta et al. 2002 argue that there should be a
strong connection between the heated gas and the activity of the central AGN.

\section{Radio Observations}
MKW~3s  was observed in August 2003 with the Giant Metrewave Radio 
Telescope (GMRT, Pune, India) at 1.28 GHz, 604 and 235 MHz. 
The 235 MHz observations suffered from major interference, and could not be 
used.
The source was observed in snapshot mode for a total of 1 hour at 1.28 GHz 
and 1.5 hours at 604 MHz, in spectral line mode (128 frequency channels) 
with a total bandwidth of 32 MHz and 16 MHz respectively.
The data reduction was carried out with the NRAO Astronomical Image 
Package Software (AIPS).
We produced a set of images with resolutions ranging from 
$\sim 5^{\prime\prime}\times5^{\prime\prime}$ to 
$\sim 17^{\prime\prime}\times16^{\prime\prime}$ at both frequencies, 
in order to highlight both the small scale features and to image the
large scale low surface brightness emission. The rms in our images is in 
the range 0.05 mJy/beam to 0.3 mJy/beam.

The left and right panels of Figure~2 show the radio contours at 1.28
GHz  overlaid to the
\chandra ~ image and temperature map of   
 MKW3s. The resolution of this radio map is $16.8^\second \times 15.8^\second$.
The radio emission from the center of MKW~3s consists of
three distinct components, i.e. a  point source associated with
the nucleus of the dominant cD galaxy NGC5920, which contains only a tiny
fraction of the total flux in the source, an extended region located south 
of the point source (southern lobe), and a very low surface brightness region 
located north--east of the compact region (northern lobe).

We notice that: i) that the southern radio lobe fills
the Souther X-ray hole; ii) both  radio lobes  seem to be contained
(confined) by (or interacting with) the extended hot regions close to the cluster center.

In Figure~3 we show  the radio contours at 604 MHz  overlaid to the temperature map of   
 MKW3s.  Unfortunatelly, due to the presence of  residual errors in the data,
 in this observation  the northern lobe is not properly imaged. 
 
Finally in Figure~4 we  show the VLA (Very Large Array, Socorro, New Mexico, USA),
A configuration 327 MHz  radio contours from Young (2004)  overlaid to the temperature map.
The   resolution of  this radio map is $\sim 5.4^{\prime\prime}\times5.4^{\prime\prime}$. 

Unlike the 604 MHz, in this lower frequency observation the northern
lobe is  well imaged. Similarly to the 1.28~GHz observation, it shows that  both
radio lobes  may be interacting with the extended hot regions.


We conclude this section by noticing that  the southern lobe is the
dominant 
feature at all frequencies, and it is
characterized by a filamentary structure, with a ridge of emission
containing most of its flux density. 
Each component in the source has a very steep spectrum, in agreement with 
the reported ultra--steep spectrum (De Breuck et al. 2000), 
 including the compact component, coincident with the 
optical galaxy, whose spectral index is 
$\alpha_{327~MHz}^{1.28~GHz} \sim 1.1$.


\begin {figure}[h]
\vskip 0cm
\centerline{\epsfxsize=.9\columnwidth \epsfbox{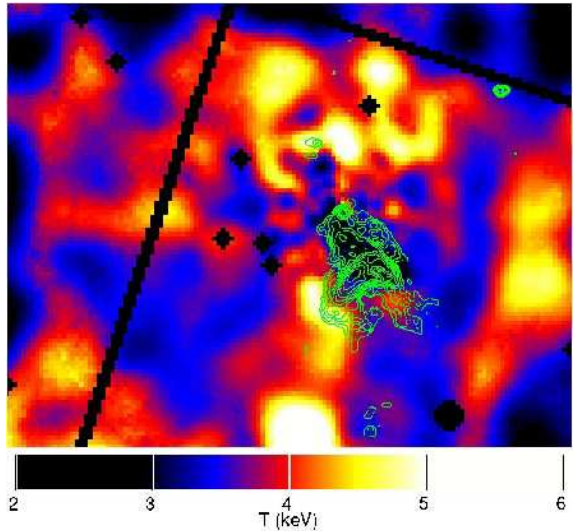}}
\vskip -0.2cm
\label{fig:tmap_610}
\caption{Temperature map of   
 MKW3s  as shown in right panel of Figure~1 overlaid to
 the~604 MHz radio contours  obtained with the GMRT.
The resolution of the radio map is $5.6^\second \times 4.6^\second$.
Isoncontours levels are  equispaced  by a factor of $\sqrt{2}$ with the
lowest contour corresponding to 0.9 mJy.}
\vskip -0.5cm
\end{figure}

\begin {figure}[h]
\vskip 0cm
\centerline{\epsfxsize=.9\columnwidth\epsfbox{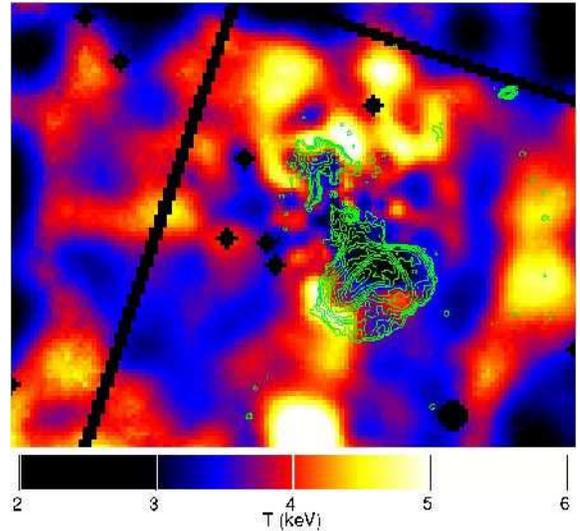}}
\vskip -0.2cm
\label{fig:tmap_VLA}
\caption{Temperature map of   
 MKW3s  as shown in right panel of Figure~1 overlaid to
 the~327  MHz radio contours  obtained with the VLA from  Young (2004).
The resolution of the radio map is $5.4^\second \times 5.4^\second$.
Isoncontours levels are  equispaced  by a factor of $\sqrt{2}$ with the
lowest contour corresponding to 2 mJy.}
\vskip -0.5cm
\end{figure}



\section{Spectrum and Age of  The radio Lobes}

We use the radio spectral informtion to estimate the radiative age of the radio
lobes as shown in Figure~5.
The spectrum of both lobes  is very steep.
The spectral index of the southern and northern lobe
between 327--604 MHz is $\alpha \sim$ 2.5 and 1.9, respectively.
The 1.28 GHz points obtained by the GMRT observations reveal 
a steepening at higher frequencies of both  components.


Assuming an injection spectrum of the relativistic
           electrons $N(\gamma)\propto \gamma^{-\delta}$ with 
           $\delta \leq 3$, a continuous injection 
model (e.g., Kardashev 1962), in which
electrons are continuously injected in a region of constant
magnetic field, is not able to reproduce the data due to
           the very steep spectrum.
At the same time, due to the combination of a steep spectrum
and of  a only  moderate curvature, also a JP model 
(Jaffe \& Perola 1973), in which a coheve population 
of relativistic electrons simply age in a region of constant 
magnetic field, is not a good model for the data.

One alternative is to use a model of the emitting plasma
in which electrons have been continuously injected in the
past and are now simply aging due to the radiative losses
(Figure~5). In this case, if the value of the magnetic field in the
two components is in the range $\sim 1-3.5 \mu$G (which is also 
barely consistent with the equipartition value) then the 
age of the radio lobes should be in the range
$\sim 0.15-0.2$ Gyrs and
the process of continuous injection should have been stopped
about $\sim 2 \times 10^7$ yrs ago.

A second possibility is that the two extended radio lobes are
buoyant bubbles injected by the central source 
about $10^8$yrs ago which are in the process of being mixed 
with the ICM.
In this case magnetized filaments are produced by the 
development of plasma instabilities in the radio volume and
the total synchrotron emission results from the convolution
of different spectra produced by relativistic electrons emitting
in regions with different values of the magnetic field strength
(e.g., Tregillis, Jones, Ryu 2004).
It is well known that the resulting convolution of the
synchrotron Kernel with magnetic
field intensity and geometry 
yields a total spectrum 
which is stretched and thus not straightforwardly related to the spectrum
of the emitting electrons 
(e.g., Eilek \& Arendt 1996; 
Katz--Stone \& Rudnick 1999) and which
depends also on the power spectrum of the magnetic
field fluctuations.

Future observations at higher frequency are required in order
to better test different modelings.

\begin {figure}[t]
\vskip 0cm
\centerline{\epsfxsize=.9\columnwidth\epsfbox{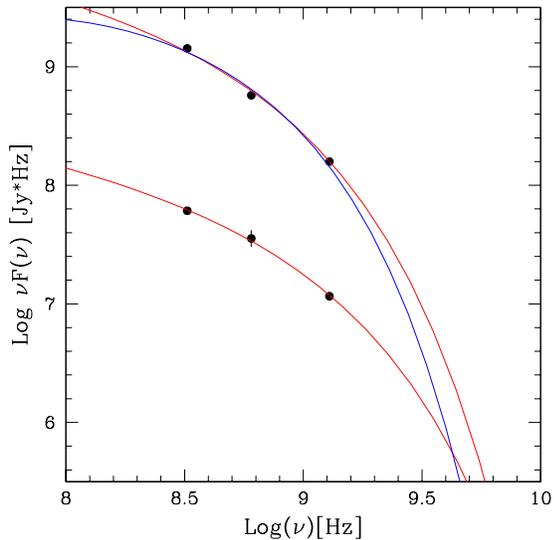}}
\vskip -0.2cm
\label{fig:6}
\caption{Data points at 327, 604 and 1287 MHz are reported for 
the southern (upper) and northern (lower) extended 
components. Relavant example of modellings are also
compared with the data.
The red curves are synchrotron models with a slope 
of the injected electrons $\delta =2.6$, and 
break and cut--off frequencies $\nu_b$=1 GHz and 
$\nu_c$=100 GHz, and
$\nu_b$=16 GHz and $\nu_c$=160 GHz for the southern
and northern components, respectively.
The blue model is obtained (for the southern component
only) with $\delta =2.0$, $\nu_b$=0.63 GHz and
$\nu_c$=25 GHz.
}
\vskip -0.5cm
\end{figure}


\begin {figure}[t]
\vskip 0cm
\centerline{\epsfxsize=.9\columnwidth\epsfbox{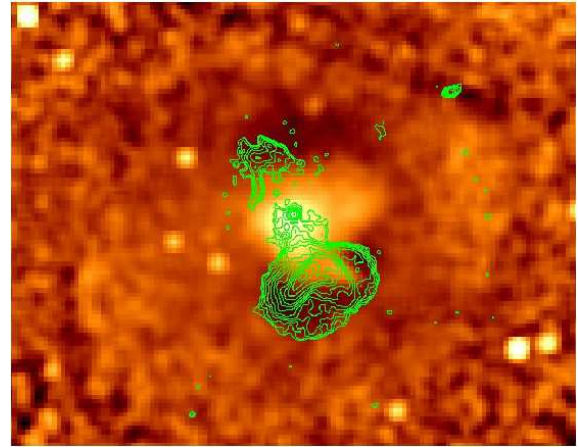}}
\vskip -0.2cm
\label{fig:5}
\caption{Unsharp-masked image of  
 MKW3s (see text)   overlaid to
 the~327~MHz VLA radio contours as shown in 
 Figure~4. 
This map indicates the presence of a second
X-ray hole opposite to the southern one filled by the  northern radio
lobe.}
\vskip -0.5cm
\end{figure}

\section{Discussion and Conclusions}

In this paper we compare the \chandra ~ X-ray observation of MKW~3s
with some radio observations obtained with the GMRT and VLA. 
The radio emission in the cluster center  is consistent with
been produced by AGN. 
Figures~2, 3, and 4 clearly show that at any  frequency there is a
good match between the hot temperature structures in the  ICM and
the shapes of the radio lobes. In particular it looks like that the hot
gas structure some how interacts and/or confines the radio emission.

It is important to say that, as reported by Mazzotta et al. (2002), the
X-ray image of MKW~3s  shows the presence of only one X-ray cavity coincident
with the souther radio lobe (see left panels of Figure~1 and Figure~2)
and there are no clear evidences for other cavities opposite to it
with respect to the X-ray peak. Nevertheless, with these new radio data
in hand, we tried to push the limit of the imaging analysis and we
manipulated the original image to produce  an  unsharp-masked image.
To do that we first smoothed the original image with a Gaussian filter with
 $\sigma=10$ and $\sigma=1$ (image ${\rm map_L}$ and ${\rm map_H}$,
 respectively) and we calculated the unsharp-masked image as 
${\rm map_L}-{\rm map_H}/{\rm map_L}+{\rm map_H}$.
In Figure~6 we report the unsharp-masked image of  
 MKW3s   overlaid to the~327~MHz VLA radio contours as shown in 
 Figure~4. 
It is interesting to note that this  map indicates the presence of a second
X-ray hole opposite to the southern one which seems to be well
correlated with the  northern radio lobe.

All the above considerations  represent a pretty good evidence that there should be a strong
connection between the shape of the radio lobes and the thermal
structure. The implications of this connection will be discussed in details in forthcoming paper.
Here we limit ourself to  a very short summary.

%
In particular we say that one possibility to explain the data 
is that the radio lobes 
       expand in the ICM and heat the gas. This would represent
the first evidence of significant
amount of gas heated  by the AGN. Nevertheless, it is not clear which
is the physical mechanism responsible for the heating. 
 Shock heating,  in fact,  can be partially excluded  as, 
in correspondences of the hot regions, there are no
 strong evidences of gas density jumps  consistent with
shock wave propagation.


\acknowledgements{We are gratefully to L.Rudnick for providing us with
  the 327~MHz VLA radio map of MKW3s. This work is supported by CXC grants GO3-4163X and by 
European contract MERG-CT-2004-510143.}

\end{document}